\documentclass{acmart}
\AtBeginDocument{%
  }

\acmConference[ISSAC '23]{International Symposium on Symbolic and Algebraic Computation}{July 24--27,
  2023}{Troms\o, Norway}
\acmPrice{15.00}
\acmISBN{978-1-4503-XXXX-X/18/06}

\usepackage{subcaption}
\usepackage[shortlabels]{enumitem}
\setlist[enumerate]{leftmargin=5mm}
\setlist[itemize]{leftmargin=5mm}
\usepackage{pgfplots}
\pgfplotsset{compat=1.18}
\usetikzlibrary{intersections}

\newcommand{\beq}{\begin{equation}}
\newcommand{\eeq}{\end{equation}}
\newcommand{\psups}[2]{\ensuremath{\left({#1}_1,\dots,{#1}_{#2}\right)}}
\newcommand{\sups}[2]{\ensuremath{{#1}_1,\dots,{#1}_{#2}}}

\newcommand{\fld}[1]{\mathbb {#1}} 
\newcommand{\cfld}[1]{\overline{\fld{#1}}} 
\newcommand{\wh}[1]{\ensuremath{\widehat{#1}}}
\newcommand{\ov}[1]{\ensuremath{\overline{#1}}}
\def\grob{\text{Gr\"obner basis }}

\newcommand{\act}{\alpha}
\newcommand{\gva}{\ensuremath{\mathcal{G}}} 
\newcommand{\g}{\ensuremath{{\bf g}}} 
\newcommand{\gp}{\ensuremath{\lambda}} 

\newcommand{\oid}{\ensuremath{O}} 
\newcommand{\ova}{\ensuremath{\mathcal{O}}} 
\newcommand{\cs}{\mathcal{K}}
\newcommand{\opsz}{\mathcal{U}}  
\newcommand{\sfu}{C^\infty\left(\opsz\right)}
\newcommand{\sfug}{\sfu^\gva}
\newcommand{\zva}{\ensuremath{\mathcal{Z}}}
\newcommand{\bin}{\ensuremath{\mathcal{B}}} 
\newcommand{\iva}{\ensuremath{\mathcal{I}}}
\newcommand{\ideal}[1]{\ensuremath{\left<#1\right>}}
\newcommand{\sig}{\ensuremath{\mathcal{S}}}
\newcommand{\inv}{\iota} 
\def\pz{{\bf z} }
\def\apz{{\tilde{\pz}} }
\def\pp{{\bf p} }
\def\app{{\tilde{\pp}} }
\newcommand{\dimz}{n}
\newcommand{\dimo}{s}

\def\wt{\widetilde}
\newcommand{\jt}[2]{{#1}^{(#2)}}
\newcommand{\td}[2]{\frac {D#1}{D#2}}
\newcommand{\pd}[2]{\frac {\partial #1}{\partial #2}}
\newcommand{\dd}[2]{\frac {d#1}{d#2}}

\newcommand{\ther}[1]{Theorem~\ref{#1}}

\newcommand{\exr}[1]{Example~\ref{#1}}
\newcommand{\defr}[1]{Definition~\ref{#1}}

\copyrightyear{2023}
\acmYear{2023}
\setcopyright{acmlicensed}\acmConference[ISSAC 2023]{International Symposium on Symbolic and Algebraic Computation 2023}{July 24--27, 2023}{Tromsø, Norway}
\acmBooktitle{International Symposium on Symbolic and Algebraic Computation 2023 (ISSAC 2023), July 24--27, 2023, Tromsø, Norway}
\acmPrice{15.00}
\acmDOI{10.1145/3597066.3597149}
\acmISBN{979-8-4007-0039-2/23/07}

\begin{document}
  \theoremstyle{acmplain}
  \newtheorem{method}[theorem]{Method}
    \theoremstyle{acmdefinition}
    \newtheorem{remark}[theorem]{Remark}
\title{Invariants: Computation and  Applications}

\author{Irina A.~Kogan}
\email{iakogan@ncsu.edu}
\orcid{0000-0001-8212-6296}
\affiliation{%
\institution{North Carolina State University}
  \streetaddress{******}
  \city{Raleigh}
  \state{North Carolina}
  \country{USA}
  \postcode{*****}
}

\begin{abstract} Invariants withstand transformations and, therefore, represent the essence of objects or phenomena. In mathematics, transformations often constitute a group action. Since the 19th century, studying the structure of various types of invariants and designing methods and algorithms to compute them remains an active area of ongoing research with an abundance of applications. In this incredibly vast topic, we focus on two particular themes displaying a fruitful interplay between the differential and algebraic invariant theories. First, we show how an algebraic adaptation of the moving frame method from differential geometry leads to a practical algorithm for computing a generating set of rational invariants. Then we discuss the notion of differential invariant signature, its role in solving equivalence problems in geometry and algebra, and some successes and challenges in designing algorithms based on this notion. 
\end{abstract}




\maketitle
\section{Introduction}\label{sect-intro}
The birth of classical invariant theory is  attributed to Arthur Cayley's 1845 paper \cite{cayley1845}, who, in his own words, was inspired by George Boole's publications  \cite{boole1841, boole1842}  on homogeneous functions under linear transformations.  Titanic  efforts of great mathematicians of the time  were devoted to the explicit computation of generating sets of polynomial invariants for homogeneous polynomials in $m$ variables (called \emph{$m$-ary forms} or \emph{quantics}) under linear changes of variables. Even in the simplest case of binary forms (polynomials in two variables) this task was overwhelming. Moreover, Cayley himself mistakenly conjectured that  there is no finite generating set of  polynomial invariants   for binary forms of degrees greater than seven \cite{cayley1855}.  Paul Gordan in 1868 \cite{gordan}, however, proved  the existence of finite generating sets  for polynomial invariants of binary forms of any degree, while Hilbert in 1890 \cite{HilbertEng},  using different techniques, extended  Gordan's  result to forms in  any  number of  variables of any degree. Hilbert's work  ended the classical period of invariant theory and stimulated  the rapid development of new fields of modern algebra, representation theory, and  algebraic geometry on whose crossroads  the algebraic invariant theory now dwells.   Excellent expositions  of  theoretical and computational aspects of modern algebraic invariant theory can be found in \cite{PV89, Sturmfels1993, DK2015}.

The birth of differential invariant theory can be attributed to Sophus Lie's work \cite{lie80, lie84} in 1880s  on infinitesimal transformations and differential invariants. Lie's ambition was to generalize Galois theory to differential equations. Arthur Tresse \cite{Tresse}, a student of Lie, proved a differential analog of Hilbert's basis theorem,  the existence of a  finite generating set  of a differential algebra of differential invariants.   \'Elie Cartan \cite{C53} based his method for solving equivalence problems   on computing relationships between differential invariants. Excellent expositions of  modern theories rooted in  Lie's and Cartan's work can be found  in \cite{olver:yellow, olver:purple, il}.

In  this tutorial, after  reviewing  basic definitions and facts about  group actions and invariants in Section~\ref{sect-intro},  we focus on two topics  lying on the intersection of algebraic and differential invariant theories.  In Section~\ref{sect-mf}, we describe the basics of the  equivariant moving frame method developed in \cite{FO99} and   show  that its algebraic interpretation leads to a practical algorithm for computing a generating set of rational invariants \cite{hk07:focm, hk07:jsc}. In Section~\ref{sect-sig}, we define the notion of differential signature  \cite{calabi98} and discuss   how it can be used to solve  equivalence problems for binary forms, as well as for smooth and algebraic planar curves. Our goal is to convey the core ideas in an  accessible and practical manner, often through examples, forgoing full generality, while providing references to more comprehensive expositions.

\section{Group actions, orbits, invariants}\label{sect-intro}
\begin{figure}
    \hspace{-1cm}
  \begin{subfigure}[b]{0.25\linewidth}
    \includegraphics[width=1.2\linewidth]{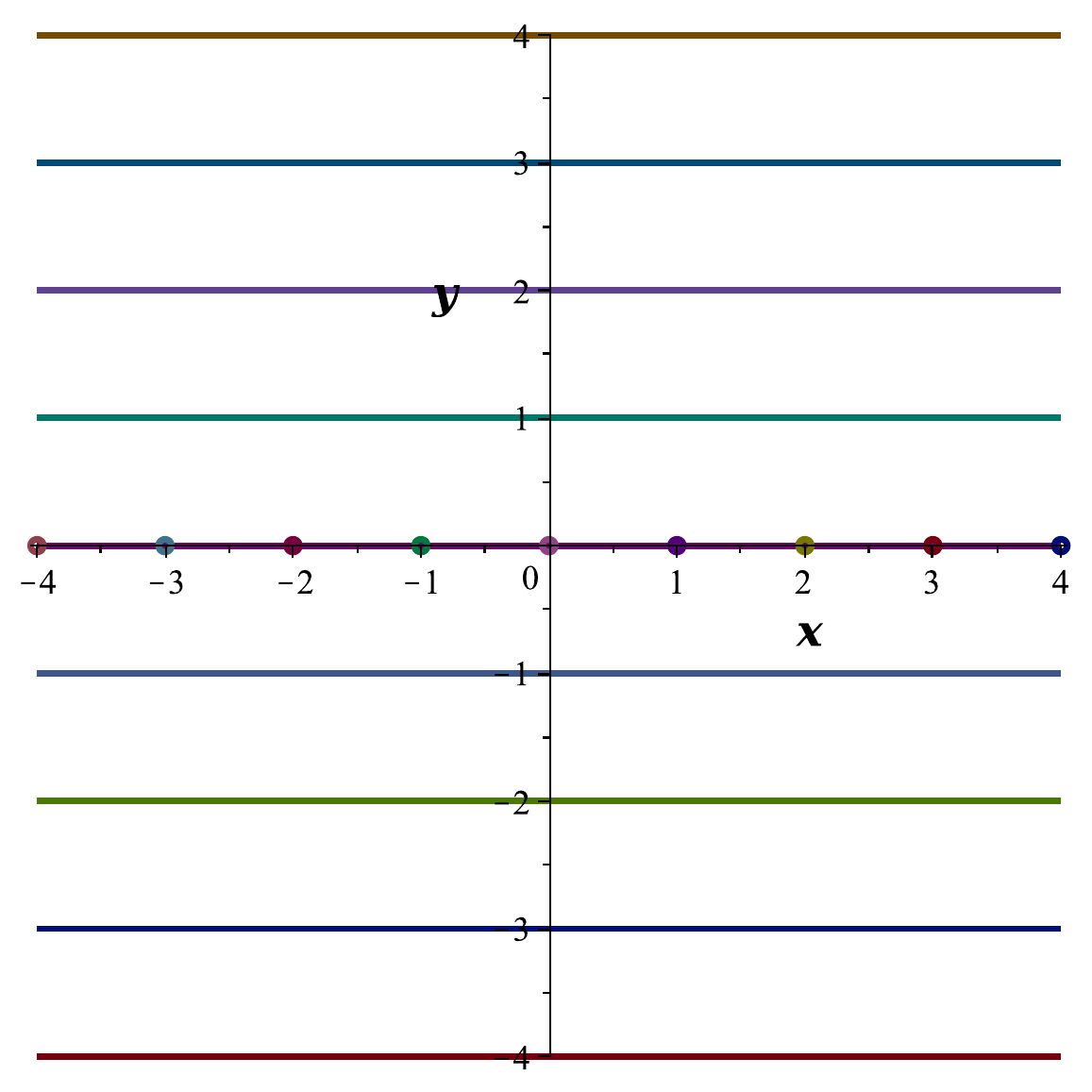}
     \caption{$\alpha_1$-orbits.}
  \end{subfigure}
  \hspace{0.4cm}
  \begin{subfigure}[b]{0.25\linewidth}
   \centering
    \includegraphics[width=1.2\linewidth]{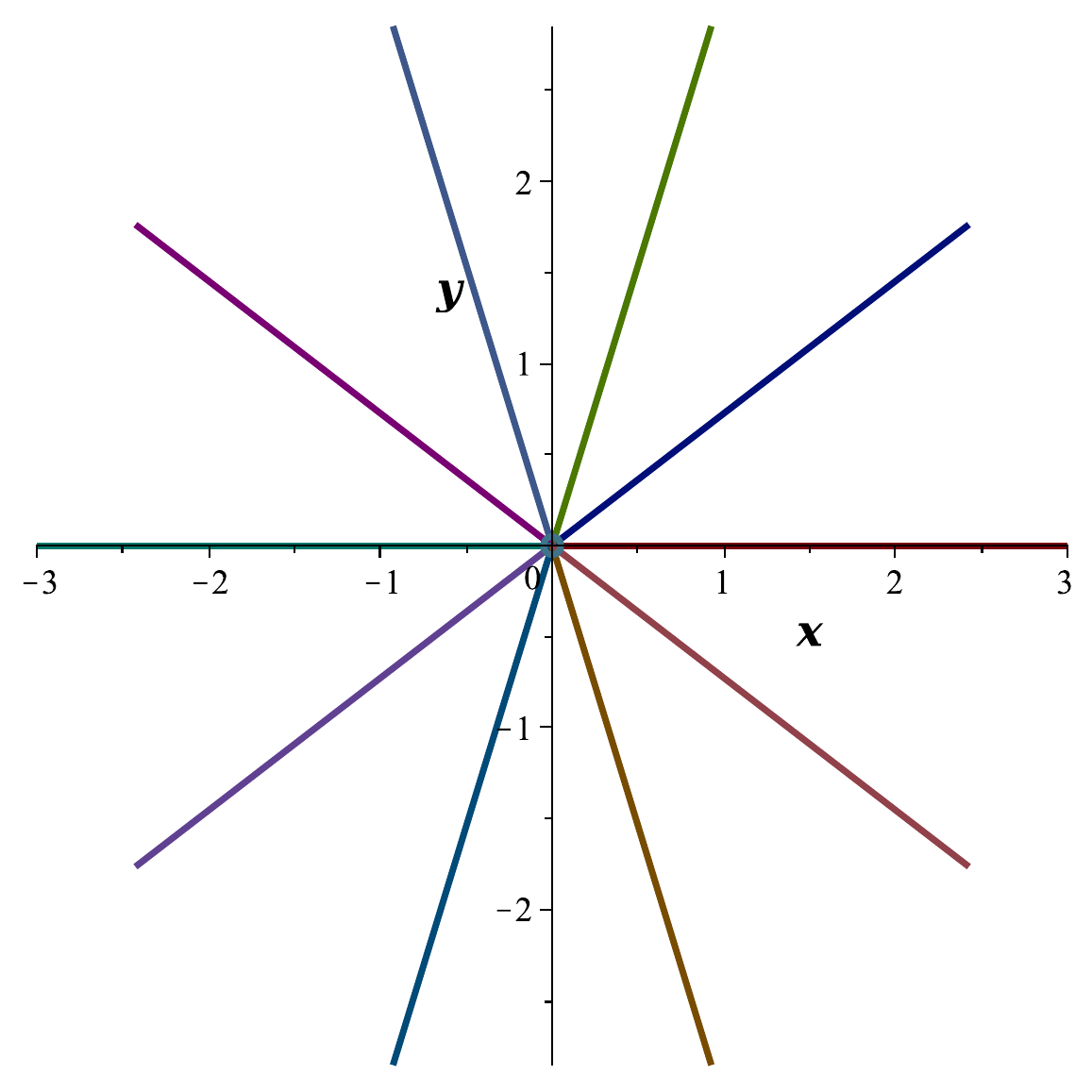}
     \caption{$\alpha_2$-orbits.}
  \end{subfigure}
\hspace{0.4cm}
  \begin{subfigure}[b]{0.25\linewidth}
   \centering
    \includegraphics[width=1.2\linewidth]{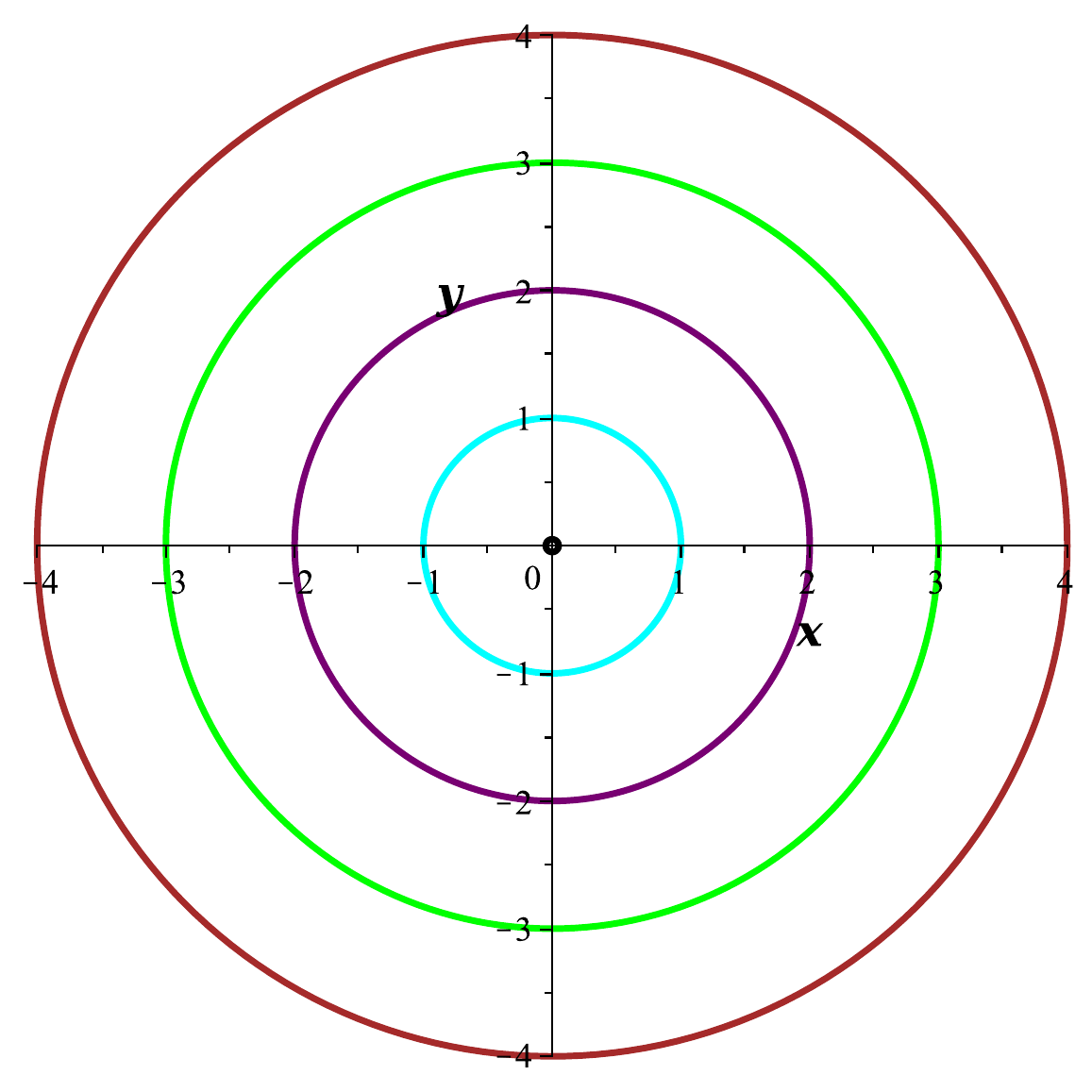}
     \caption{$\alpha_3$-orbits.}
  \end{subfigure}
\caption{{Orbits under the actions in \exr{ex-act-orb}.}}
 \Description{Three pictures. On the first one, several horizontal lines and dots along the x-axis. On the second - open rays originated at the origin. On the third - concentric circles centered at the origin.}
\label{fig:orb}
 \end{figure}

We start with standard and simplest definitions, and then  make adjustments necessary  to work with local smooth  and rational actions. 
\begin{definition}[action]\label{def-act}
An  \emph{action}  of a group $\gva$  on a set $\zva$ is a map $\act\colon  \gva \times \zva \to \zva $  satisfying the following properties\footnote{Here, $\cdot$ denotes the  group operation. The abbreviation  $\alpha(\g,\pz)=\g\pz$ is often used in the literature and  we use it in Figure~\ref{fig-rho}.}:

\begin{enumerate}
    \item[i.] \label{ai} \emph{Associativity:}  $ \act(\g_1 \cdot  \g_2, \pz) = \act(\g_1, \act (\g_2, \pz ))$, $\forall \g_1, \g_2\in \gva$  and  $\forall \pz\in \zva$.
\item[ii.]\label{aii} \emph{Action of the identity element:} 	 $\act(e,\pz) = \pz$, $\forall  \pz \in \zva$.  \end{enumerate}

\end{definition}
\begin{definition}[orbit]\label{def-orb}  The \emph{orbit} of $\pz\in\zva$ under an action $\alpha$ is a set
\beq\label{ovaz}\ova_{\pz}=\left\{\apz\in \zva|\exists \g\in\gva:\, \apz=\alpha(\g,\pz)\right\}.\eeq
An  action is called  \emph{transitive} if  $\ova_{\pz}=\zva$ for some (and, therefore, for all) $\pz$. 
\end{definition}
Since  orbits partition $\zva$ into equivalence classes, many  classification problems  can be restated as problems of orbit classification.  
\begin{example}\label{ex-act-orb}For $\gva =\fld R=\{t\}$, with the addition as the group operation, and $\zva={\fld R}^2=\{(x,y)\}$, consider four actions:
\begin{align}
\label{a1} \act_1(t,(x,y)) &=\left(x +ty, \, y\right),\\
 \label{a2}\act_2(t,(x,y)) &=\left(e^t \,x, \, e^ t \, y\right),\\
\label{a3} \act_3\left(t,(x,y)\right)&=\left(x\cos t-y\sin t, \, x\sin t+y\cos t\right),\\
\label{a4} \act_4(t,(x,y)) &=\left(e^t \,\left(x\cos t-y\sin t\right), \, e^ t \, \left( x\sin t+y\cos t\right)\right) 
 \end{align}
Under $\alpha_1$, the orbit of any point on the $x$-axis  is the  point itself and the orbit of any other point is the horizontal line through the point.  Under  $\alpha_2$, the orbit of the origin is the origin itself and the orbit of any other point is the open ray, whose vertex is at the origin, passing through the point.  Under $\alpha_3$,  the orbit of the origin  is the  origin itself and  the orbit of any other point is the circle centered at the origin  passing through the point. Figure~\ref{fig:orb} shows the orbits under these three actions. Drawing the orbits under $\alpha_4$ is left for the reader.
 \end{example}
\begin{definition}[stabilizer]\label{def-stab} The \emph{stabilizer} of $\pz\in\zva$ under an action $\alpha$ is the set $\gva_{\pz}=\{\g\in \gva\,|\, \alpha(\g,\pz)=\pz\}$. The action is called \emph{free} if the stabilizer of every point is $\{e\}$.
\end{definition}
It is easy to show that $\gva_{\pz}$ is a subgroup\footnote{An alternative terms  used for the stabilizer is the \emph{isotropy group}  of $\pz$.} 
of $\gva$. The intersection $\gva_{\zva}=\cap_{\pz\in \zva} \gva_{\pz}$ is a normal subgroup of $\gva$, called the  \emph{global stabilizer}. The induced action  of the quotient group $\gva\backslash \mathcal \gva_{\zva}$ has the same orbits as the action of $\gva$  and trivial global stabilizer. Actions with trivial global stabilizer are called \emph{effective}.
\begin{example} None of the actions in \exr{ex-act-orb} are free. 
For the action $\alpha_3$, the stabilizer of the origin is $\fld R$, while the stabilizer of any other point is  a proper  subgroup  $\mathcal H=\{2\pi n\, |\, n\in \fld Z\}$, which is, therefore, the global stabilizer.  The quotient $\gva\backslash \mathcal H\cong SO_2(\fld R)$ is the special orthogonal group.  The induced action of $SO_2(\fld R)$ is  free away from the origin.
\end{example}
\begin{definition}[invariants]\label{def-inv} A function $f$ with the domain $\zva$ is called invariant\footnote{Although ``invariant'' here is used as an adjective, it is also often used as a noun. For instance, the  term ``rational invariant'' is often used instead of  ``rational invariant function''.} under an action $\act$ if it is constant along each orbit:
\beq\label{invf}f\left(\alpha (\g,\pz)\right)=f(\pz), \qquad \forall (\g,\pz)\in \gva\times\zva. \eeq
\end{definition}
A constant function on $\zva$ is trivially an invariant for any action, but it is not useful for describing the orbit-space. Ideally, we would like to find an invariant function, or a set of invariant functions,  that takes distinct values  on distinct orbits. Such set is called  \emph{separating}. More, generally: 
\begin{definition}[separating set]\label{def-sep} Assume $\{\zva_\tau\,|\, \tau\in \mathcal T \}$, where $\mathcal T$ is some index set, is a partition of a set $\zva$:
$$\zva=\displaystyle{\cup_{\tau\in \mathcal T} \zva_\tau} \text{ and  } \zva_{\tau_1}\cap \zva_{\tau_2}=\emptyset, \text{ when } \tau_1\neq\tau_2.$$  A subset $\mathcal F$ of functions on $\zva$ \emph{separates the partition} if for any $\pz_1,\pz_2\in \zva$:
\begin{align*} \exists f\in \mathcal F, \text{ s.t. }f( \pz_1)\neq  f(\pz_2)& \Longleftrightarrow  \\&\exists \tau_1\neq \tau_2\in \mathcal{T} \text{ s.t. } \pz_1\in \zva_{\tau_1} \text{ and }\pz_2\in \zva_{\tau_2}.
\end{align*}
\end{definition}

\begin{example}\label{ex-inv}
In \exr{ex-act-orb},  for the action $\alpha_1$, an invariant function $v(x,y)=y$ separates all one-dimensional orbits, but not the zero-dimensional orbits along the $x$-axis. For the action $\alpha_2$,  there are no non-constant continuous invariants. A function $f(x,y)=\frac{x}{\sqrt{x^2+y^2}}$ is invariant, but it is undefined at the origin and it does not separate all one-dimensional orbits, unless it is joined by another invariant,  $g(x,y)=\frac{y}{\sqrt{x^2+y^2}}$.\footnote{We invite the reader to find two different orbits on which $f$ takes the same value, and explain how $g$ helps to separate them.}   There is a simple rational invariant  $h(x,y)=\frac{y}{x}$ with even smaller domain, which is also not separating.  For  $\alpha_3$, $r(x,y)=\sqrt{x^2+y^2}$
 is a separating invariant.  For $\alpha_4$, a set of continuous separating invariants \eqref{inv-sp} on ${\fld R}^2\backslash \{(0,0)\}$ is derived in \exr{ex-cs}.
 \end{example}
If $\gva$ and $\zva$ are topological spaces, it is natural to require the action $\alpha$  and the invariants to be \emph{continuous} maps,  however,   it is also often  useful to restrict their domains. 
\begin{definition}[Local] \label{def-local}\hfill
\begin{itemize}
\item We define a \emph{local action} by allowing the domain of $\act$ in  \defr{def-act} to be an open subset $\Omega$ of $\gva\times \zva$, such that   $\{e\}\times \zva\subset\Omega$ and requiring property (i) to hold whenever $(\g_2,\pz)$ and $(\g_1\cdot\g_2,\pz)$ are in $\Omega$. 
 A \emph{local action is  free} if  
$$\alpha(\g,\pz)=\pz \Longrightarrow \g=e\text{ for  any }(\g,\pz)\in \Omega.$$
\item We define a \emph{local invariant function} by allowing its domain in \defr{def-inv} to be an open subset $\mathcal U \subset \zva$ (locality of the domain) or  allowing \eqref{invf} to hold only for  all $(\g,\pz)\in \mathcal W$, where $\mathcal W\subset \Omega$ is an open subset, such that   $\{e\}\times \mathcal U \subset\mathcal W$ (locality of the invariance). 
\end{itemize}
\end{definition}

In what follows we will consider two scenarios\footnote{Both scenarios have natural generalizations, but we give the simplest possible setup to convey the main ideas.}:
\begin{enumerate}
    \item[i.]  \emph{Smooth:} $\gva$ is a real Lie group, $\zva$ is an open subset of ${\fld R}^n$ with metric topology, and $\act$ is a smooth map.   
    
    \item[ii.]\emph{Algebraic:} 	$\gva$ is an affine algebraic group of positive dimension over a field ${\fld K}$ of characteristic zero, $\zva={\fld K}^n$ with Zariski topology, and   $\act$ is a rational map. \footnote{More precisely, we consider the  restriction of a rational action over $\cfld K$.}  
     \end{enumerate} 

\section{Cross-section and invariantization} \label{sect-mf}
We present here the \emph{cross-section}  method,  introduced in  \cite{FO99}, for constructing local smooth invariants and show how  its algebraic formulation  \cite{hk07:focm,hk07:jsc} leads to an algorithm for constructing a generating set of rational invariants.  

\subsection{Smooth setting}\label{sect-cs-sm}
 \begin{definition}[Smooth cross-section] \label{def-cs}A submanifold $\cs  \subset \zva$ is called  \emph{a local cross-section}  for a local  group action $\alpha\colon\Omega \to \zva$ if there exists an open subset $\opsz\subset \zva$, called the \emph{domain of the cross-section}, such that, for each $\pz \in \opsz$,  the connected component $\ova^0_{\pz}$ of  $\ova_{\pz}\cap\,\opsz$ that contains $\pz$  intersects $\cs$ transversally  at a single point:   
 $$\ova^0_\pz \cap \cs = \{\apz\}\text{ and  }T\cs|_{ \apz}\oplus T\ova_{\pz}|_{\apz}=T\zva|_{ \apz}.$$
 \end{definition}
 
Let $\dimo$ denote the maximal orbit dimension for a $\gva$-action on $\zva$. If a point $\pz_0$ belongs to an orbit of dimension $\dimo$, then the Frobenius Theorem  implies the existence of a local cross-section $\cs$ containing $\pz_0$ on some domain $\opsz$, such that $\cs$ is the zero set of a set $K$ of $\dimo$ smooth independent\footnote{Here \emph{independent} means \emph{functionally independent}.  A set of smooth functions on $\opsz$  is called independent on an open subset $\opsz$ if the set of  their differentials is linearly independent at each point of  $\opsz$, except perhaps a discrete set of points. See \cite[p.~461]{hk07:focm} and   \cite[p.~85]{olver:yellow} for details.}  functions on $\opsz$ (see \cite[Theorem 1.6]{hk07:focm}). 
  Let  $\sfu$ denote the algebra, over  $\fld R$,  of all smooth real-valued functions  on $\opsz$, and $\sfug$ the subalgebra of all locally  $\gva$-invariant functions.  %
A cross-section $\cs$ serves to define an \emph{ invariantization map} $\inv\colon \sfu \to \sfug$,
which sends a function $f$  on $\opsz$ to the unique \emph{locally} invariant function $\inv f:=\inv(f)$ that has the same values on the cross-section as $f$, i.e.~$\inv f|_\cs=f|_\cs$.  For any $\pz\in \opsz$,
$$\inv f(\pz)=f(\apz)$$ { where } $\apz=\ova^0_\pz \cap \cs$ (see \defr{def-cs}).  Obviously, $\iota$ respects all algebraic  operations  and so a choice of a cross-section  induces an algebra morphism $\inv \colon \sfu \to \sfug$, which is a projection, because $\iota f=f$  if and only if $f\in \sfug$.

Let $z = \psups z \dimz$ be coordinate functions on  the domain  $\opsz$ of the cross-section $\cs$. The $\cs$-invariantization  $ \inv z=\psups {\inv z} \dimz$  of $z$ leads to $\dimz$ invariant functions, called  \emph{normalized} invariants in \cite{FO99}. Their explicit formulas can be  obtained by expressing the coordinates  of the intersection point  $\apz=\ova^0_{\pz} \cap \cs$  (see \defr{def-cs}) in terms of coordinates of $\pz$ (see \exr{ex-cs} below). \emph{The set of  normalized  invariants separates connected components of the orbits  on $\opsz$} and   have the crucial \emph{replacement property}:  for any $f\in\sfu$, its invariantization $\inv f$  can be obtained by replacing $z$ with   $\inv z$:
\beq\label{eq-repl}\inv f(z)=f(\inv z)\eeq

Let $\cs$ be the zero set of $\dimo$ independent functions $K_1(z),\dots, K_\dimo(z)$. Then $\inv(K_i)=0, \, i=1,\dots,\dimo$, and so \eqref{eq-repl}  implies  that  
\beq\nonumber \label{fund-syz}\{K_i(\iota z)=0\,|\, i=1,\dots, \dimo\}\eeq is a  set of independent   generating functional relationships (\emph{syzygies}) among invariants  $\psups {\iota z} \dimz$. The implicit function theorem implies that, after perhaps shrinking $\opsz$,  there are $\dimz-\dimo$ functionally independent invariants $\sups {\inv z_i} {\dimz-\dimo}$  among $\inv z$, such that any other smooth local invariant on $\opsz$ is a smooth  function of those.  In other words,  $\sups {\inv z_i} {\dimz-\dimo}$ is a minimal set of smooth generators of $ \sfug$.\footnote{Minimal smooth generating sets are called \emph{fundamental}.}   
\begin{example}\label{ex-cs} Let us revisit actions and invariants of  Examples~\ref{ex-act-orb} and~\ref{ex-inv}.  For $\alpha_2$, the unit circle $\cs=\{(x,y)\,|\, x^2+y^2=1\}$ is a cross-section with the domain $\opsz=\fld R^2\backslash\{(0,0)\}$. The orbit of $(x,y)\in\opsz$ intersects $\cs$ at the point $\left(\frac{x}{\sqrt{x^2+y^2}},\frac{y}{\sqrt{x^2+y^2}}\right)=(\inv x, \inv y)$. We recognize invariants $f$ and $g$ from \exr{ex-inv}. As it is discussed above, the defining equation for $\cs$ provides us with the fundamental syzygy:  $(\inv x)^2+(\inv y)^2=1$. We invite the  reader to verify that invariantization $\tilde\iota$ defined by a vertical line cross-section $\wt\cs=\{(1,y)\}$  with the domain  $\widetilde \opsz=\{(x,y)|x>0\}$ produces normalized invariants
$(\tilde\inv x, \tilde \inv y)=(1, \frac y x)$. In agreement with the \emph{replacement property}: 
$$\tilde\inv y=\frac {\inv y}{\inv x},\,\, \inv x=\frac{\tilde\inv x}{\sqrt{(\tilde\inv x)^2+(\tilde\inv y)^2}} \text{ and } \inv y=\frac{\tilde\inv y}{\sqrt{(\tilde\inv x)^2+(\tilde\inv y)^2}}.$$

For the actions $\alpha_1$ and $\alpha_3$, we leave it to the reader to come up with cross-sections whose corresponding sets of normalized invariants contain  invariant functions listed in \exr{ex-inv}.

For the action $\alpha_4$, the unit circle $\cs$  also serves as a cross-section with the domain $\opsz=\fld R^2\backslash\{(0,0)\}$.  Since the coordinates of the intersection point $\apz=\ova_\pz \cap \cs$  are not obvious in this case, we find the group-parameter $t$ that sends a point with coordinates $(x,y)$ to $\cs$. In other words, we solve for $t$ the equation
\beq\label{eq-K-sp}\left(e^t \left(x\cos t-y\sin t\right)\right)^2+\left(e^ t \left( x\sin t+y\cos t\right)\right)^2-1=0\eeq
obtaining:
\beq\label{sp-rho}t=\rho(x,y)=-\ln r, \text {where $r=\sqrt{x^2+y^2}$}.\eeq
Substitution of \eqref{sp-rho} into \eqref{a4}  produces a set of normalized invariants:
\begin{align}\label{inv-sp}\inv x&= \frac 1 r \left(x\cos(\ln r)+y\sin(\ln r) \right), \\
 \nonumber \inv y&= \frac 1 r \left(y\cos(\ln r)-x\sin(\ln r) \right)
\end{align} 
with the fundamental syzygy  $(\inv x)^2+(\inv y)^2=1$, determined by the  defining equation of $\cs$. This is a pair of global invariants  on $\opsz$, separating the orbits. 

Alternatively, for $\act_4$, one can choose  the positive part of the $y$-axis as   a local cross-section whose domain is  the upper half-plane $\mathcal H$. It gives rise to  a single non-constant local normalized  invariant $\tilde\inv y=re^{\arctan\left(\frac x y\right)}$, which separates connected components of  the intersections of the orbits with $\mathcal H.$  
\end{example}
We are now ready to get away from various actions of $\fld R$ on  $\fld R^2$.

\begin{example}[Rotation in $\fld R^3$]\label{ex-so3} Let $SO_3$ act on ${\fld R}^3=\{(x,y,z)\}$ by rotations. The orbits are two dimensional spheres, except for the origin, which is an orbit by itself.  In this example, the maximal  orbit-dimension $\dimo=2<\dim(SO_3)=3$  and  the stabilizers of  all points away from the origin are isomorphic to $SO_2$.  The positive part of the $z$-axis serves as a cross-section on  $\opsz={\fld R}^3\backslash\{(0,0,0)\}$. The normalized invariants relative to this cross-section are
$$\inv x=\inv y=0 \text{ and } \inv z=\sqrt{x^2+y^2+z^2}.$$  
\end{example}
\emph{Differential invariants} naturally arise in the study of  the geometric equivalence problems. We restrict our attention to the simplest  case of planar curves. An  action $\act(\g,x,y)=(X,Y)$ of $\gva$ on $\fld R^2$ induces an action on  the infinite dimensional set of planar curves. Remarkably,  the equivalence problem for curves can be solved by   prolonging the action of $\gva$ to a finite dimensional jet space $J^k$ of sufficiently high order. For our purposes, we can ignore the points where a curve has  vertical tangent\footnote{By ignoring these points we restrict ourselves to a dense open subset of what  in \cite{olver:yellow} is called an extended jet bundle. See   \cite[Chapter 3]{olver:yellow}  and in particular Definition~3.27 and Example~3.29 for a detailed discussion of the relationship between  a local jet space and a global jet bundle of curves.}  and identify 
$J^k$ with $\fld R^{k+2}$. The coordinate functions on  $J^k$ are denoted by  $\left(x,y,y^{(1)},\hdots , y^{(k)}\right)$.
Although formally,  $y^{(k)}$ is  viewed as  an independent coordinate function, we define the prolongation   of the action $\act$  keeping in mind  that 
  $y^{(k)}$ is the ``place holder'' for the $k$-th derivative  of $y$ with respect to $x$:
\beq\label{pr-act} \jt Y 1=\td Y X, \quad \jt Y {\ell+1}=\td{ \jt Y {\ell}} X \text{ for } \ell=1,\dots, k-1,\eeq
where differential operator  $D=\pd{}{x}+\jt y 1 \pd{}{y}+\jt y 2\pd {}{\jt y 1 }+\dots$ is the total derivative with respect to $x$. 
Invariants under the prolonged action on $J^k$ are called \emph{differential invariants}. 
\begin{example}[Euclidean curvature] \label{ex-se2} Consider  an  action $\alpha$ of the special Euclidean group $SE_2=SO_2\ltimes {\fld R}^2$ on ${\fld R}^2$, which maps  a group element $\g$ with parameters $(c, s, a, b)$, such that $c^2+s^2=1$, and a point $\pp\in {\fld R }^2$    with coordinates $(x,y)$   to a point $\app\in {\fld R }^2$with coordinates $(X,Y)$:
\beq\label{se2-act}\begin{bmatrix}X\\Y\end{bmatrix}= \begin{bmatrix}c &-s\\s &c\end{bmatrix}\begin{bmatrix}x\\y\end{bmatrix} +\begin{bmatrix}a\\b\end{bmatrix}.\eeq
This action is the composition of the rotation by an angle $t$, such that  $\cos t =c$ and $\sin t =s$ and the translation by a vector $(a,b)$. 
 The action is transitive, and, therefore, has no non-constant invariants. However, its prolongation to   $J^2$,  defined by \eqref{pr-act}, 
\beq\label{se2-j2}  \jt Y 1=\frac {s+c\jt y 1 }{c-s \jt y1 }, \quad \jt Y 2=\frac{ \jt y 2} {({c-s \jt y1})^3}\eeq
 is a local (in fact, a rational) non-transitive action with the domain
 $$\Omega:=\{(\g,\pz)\,| \,c-s\jt y1 \neq 0\}.$$
It follows that generic  orbits $\ova_{\pz}$ have two components, and on each of these components  the sign of the last coordinate,  $\jt y 2$, of $\pz\in J^2$ is preserved. 
 The $\jt y 2$-coordinate line  in $J^2$, defined by equations:
  \beq\label{cs-se2}x=0, \quad y=0,\quad \jt y 1 =0\eeq is a cross-section for the prolonged action \eqref{se2-act} - \eqref{se2-j2}.  To find a group element  that brings a point $\pz\in J^2$ to the cross-section,  we start by solving the equations $X=Y=\jt Y 1=0$ for $c, s, a$ and $b$:
\begin{align}\label{se2-rho}c=\pm \frac 1 {\sqrt{\left(\jt y 1\right)^2+1}},\,\,  s=\mp\frac {\jt y1} {\sqrt{\left(\jt y 1\right)^2+1}} \\
\nonumber  a=\frac{ -y\jt y 1-x}{\sqrt{\left(\jt y 1\right)^2+1}} ,\, b=\frac{ x\jt y 1-y}{\sqrt{\left(\jt y 1\right)^2+1}}.
\end{align}
The freedom of choosing $\pm$ or $\mp$  signs in \eqref{se2-rho} is a reflection of  the fact  that  the two connected components of $\ova_\pz$ intersect $\cs$ at two distinct points (unless $\jt y2(\pz) =0$).   The intersection point of the connected component $\ova_\pz$ containing  ${\pz}$ is obtained when 
\beq\label{se2-ineq}c-s\jt y 1>0 \eeq and so, in \eqref{se2-rho}, ``$+$''  must be chosen     in the first formula  and ``$-$'' in the second. Substitution of \eqref{se2-rho}, with this choice of signs,  into the second formula of  \eqref{se2-j2} produces the only non-constant  normalized invariant
\beq\label{kappa} \inv \jt y 2= \frac {\jt y2} {\left(\left(\jt y 1\right)^2+1\right)^{\frac 3 2}}=\frac {y_{xx}} {\left(y_x^2+1\right)^{\frac 3 2}} ,\eeq
in which, after replacing jet variables with the corresponding derivatives, we recognize the Euclidean signed curvature $\kappa$ of a curve given as a graph $y=\varphi(x)$.  This is an example of a local invariant,  in a sense of ``locality of invariance'' per \defr{def-local}. Indeed, $\kappa$  remains invariant when the prolonged  action $\alpha$, defined by \eqref{se2-act} - \eqref{se2-j2}, is restricted to a smaller domain $ \mathcal W=\{(\g,\pz)\,| \,c-s\jt y 1>0\}\subset\Omega$, but takes two opposite in sign values on each connected components  of an orbit  when $\alpha$ is defined on its full domain $\Omega$.   
 \end{example}
In the above example, as well as when we considered action $\alpha_4$ in \exr{ex-cs}, explicit expressions for normalized invariants
were far from obvious and computed in the following steps:
\begin{method}\emph{({Normalized invariants via smooth cross-section})}\label{sm-method}
Let $\act: \Omega\to \zva$ be a local action of $\gva$ on $\zva$, with the domain $\Omega$, such that $\{e\}\times \zva\subset \Omega\subset \gva\times \zva$.    
\begin{enumerate}[i.]
\vskip1mm
\item\label{st-cs}\emph{(Choose a cross-section)} On an open subset $\opsz\subset\zva$, such that orbits of all points in $\opsz$ are of maximal dimension $\dimo$, choose  a (local) cross-section $\cs$ defined as the zero set of a set $K$ of $\dimo$ smooth independent functions   on $\opsz$.\footnote{E.g.~$K=\{x^2+y^2-1\}$  for $\alpha_4$  in   \exr{ex-cs} and  $K=\{x,y, \jt y1\}$ in \exr{ex-se2}.} 

\item\label{st-eq}\emph{(Substitute the action)}  Substitute the coordinates of $\alpha(g,\pz)$  in $K$, obtaining $s$ smooth functions $K(\alpha(g,\pz))$ on  an appropriate open connected subset $\mathcal W\subset \Omega$, such that $\{e\}\times \opsz\subset\mathcal W$.  \footnote{See  the left-hand side of \eqref{eq-K-sp} in  \exr{ex-cs} and the right-hand sides of  $\{X, Y, \jt Y 1\}$ in formulas \eqref{se2-act} - \eqref{se2-j2} of  \exr{ex-se2}.}

\item\label{st-solve}\emph{(Solve for the group  parameters)} Find a smooth map $\rho\colon\opsz\to \gva$,  pictured in Figure~\ref{fig-rho}, such that:\footnote{In \exr{ex-cs}, the map $\rho$ is given by \eqref{sp-rho} and is obtained by solving \eqref{eq-K-sp} for $t$. In \exr{ex-se2}, the map $\rho$ is given by \eqref{se2-rho} with the top choice of signs, and is obtained by solving $\{X=0, Y=0, \jt Y 1=0\}$ for $c,s,a,b$ with an additional condition $c-s\jt y 1>0$. Note that only this choice of signs insures that \eqref{rho1} holds.}
\begin{align}
\label{rho1} \forall \pz\in \cs:\quad &\rho(\pz)=e,\\
\label{rho2}  \forall \pz\in \opsz:\quad
& \left(\rho(\pz),\pz \right)\in \text{ zero set of }K\left(\alpha(g,\pz)\right).
 \end{align}
\item \label{st-subs} \emph{(Substitute into   the action)}  Compute, in coordinates,  $\alpha(\rho(\pz), \pz)$.  Note that the previous  steps insure that $\alpha(\rho(\pz), \pz)$ is \emph{the unique point of the intersection} of the connected component $\ova^0_{\pz}$,  containing $\pz$, of  $\ova_\pz\cap \opsz$  with $\cs$, and therefore, by definition,  coordinates of  $\alpha(\rho(\pz), \pz)$
 are normalized invariants.\footnote{See  \eqref{inv-sp} in \exr{ex-cs}, while in  \exr{ex-se2}, see  \eqref{kappa}. }
 \end{enumerate}
 \end{method}

\noindent Constructability and practicality of steps i.~and iii.~needs further discussion:\\
\emph{Step i:} As noted under \defr{def-cs}, for every  point $\pz$ whose orbit is of maximal dimension $\dimo$,   there  exists  a local  cross-section, passing through this point, which is the zero set   of $\dimo$ smooth functions. However,  can we construct  a  set $K$ of such functions in an algorithmic way? If yes,  is there a preferable cross-section? The answer to the first question is yes: 
a generic manifold of codimension $\dimo$  passing through $\pz$ is a cross-section on some open domain  $\opsz\ni\pz$.  Moreover, we can always find a \emph{coordinate cross-section} whose defining equations,  $x_{i_1}=c_{i_1}, \dots, x_{i_\dimo}=c_{i_\dimo}$, are obtained by setting  $\dimo$ coordinate functions equal to constants.\footnote{ In fact,  one can  identify appropriate coordinate functions and constants by writing out the  matrix of infinitesimal generators of the group action as described in the proof of   \cite[Theorem 1.6]{hk07:focm}.}  There is no uniform answer  for the second question. Since the domain of normalized invariants equals to the domain of the cross-section $\opsz$ , one may prefer to find a cross-section whose domain is dense  \emph{dense} in $\zva$. However, for some actions this is impossible.
   Even if such a cross-section exists,  we may prefer a coordinate cross-section with a smaller domain because all non-constant normalized invariants it produces are functionally independent, and thus  comprise a fundamental set.\footnote{We invite the reader to revisit actions $\alpha_2$ and $\alpha_4$ in \exr{ex-act-orb} and, for each of the two actions,  appreciate the difference between the sets of normalized invariants obtained with the unit circle  cross-section versus a coordinate cross-section.}  In addition, the context of the problem (geometric, physical, etc.) may inform an appropriate choice of  a cross-section.
  \\
 \emph{Step iii:} If the maximal dimension of the orbits equals the dimension of the group (such actions are called \emph{locally free}), the existence of the smooth map $\rho$ is guaranteed by the transversality condition and the implicit function theorem (upon, perhaps, shrinking  the domain $\opsz$). This sufficient condition appears in \cite{FO99}.\footnote{In the absence of local freeness, the map $\rho$ can be often constructed.  For instance, for the $SO_3$-action and cross-section  in \exr{ex-so3}, the map $\rho$ is constructed in  \cite[Example 3.9]{kogan03}. See also a related notion of partial moving frames in \cite{O-rec11}.}  However, it is well known that not every system of equations can be solved in terms of elementary functions, so from  the practical perspective this step maybe impossible to preform. Even when it is possible, it can be impractical for groups of large dimensions. This last consideration stimulated  the development  of various inductive and recursive  constructions \cite{kogan01,kogan03,O-rec11,OV-rec18}.   
 \begin{figure}
\begin{minipage}[t]{0.45\linewidth}
   \begin{tikzpicture}
  \begin{axis}[
    axis lines = none,
    xmin = -3, xmax = 2,
    ymin = -1, ymax = 4,
    domain = -3:2,
    restrict y to domain = -1:4,
    width = 2.1in,
    samples = 50,
    clip=false,
    ]
    \foreach\z in {0,...,2}
    \addplot[purple, thick,
    ]
    {-0.1*x^2+\z};
    \addplot[purple, thick,
    name path = curve,
    ]
    {-0.1*x^2+3}
    coordinate[pos = .2] (p1)
    coordinate[pos = .4] (p2)
    node[left, pos=0] {${\mathcal O}_{\bf z}$};
    ;

    \addplot[
    blue, thick,
    name path = blueline,
    ]
    {-9*(x-1)}
    node [right] {$\mathcal K$}
    ;

    \path [name intersections = {of = curve and blueline, by={pt}}];

    \begin{scope}[every node/.style={circle, fill, inner sep=1pt}]
      \node (p1n) at (p1) [label={below:$\g\pz$}] {};
      \node (p2n) at (p2) [label={below:$\pz$}] {};
      \node (ptn) at (pt) [label={below left:$\tilde \pz$}] {};
    \end{scope}

    \draw [->, >=stealth, bend left=90] (p1n) to node[above] {$\rho(\g\pz)$} (ptn);
    \draw [->, >=stealth, bend left] (p2n) to node[pos=.4, above=-2.8pt] {$\rho(\pz)$} (ptn);
\end{axis}
\end{tikzpicture}
       \caption{A moving frame map $\rho$.}
    \label{fig-rho}
  \end{minipage}
\hspace{0.6cm}
  \begin{minipage}[t]{0.45\linewidth}
  \centering
   \begin{tikzpicture}[>=stealth]
  \begin{axis}[
    axis lines = center,
    axis equal,
    xlabel = $x$,
    ylabel =  $y$,
    xmin = -1, xmax = 5,
    ymin = -.5, ymax = 7,
    domain = -1:5,
    restrict y to domain = -.5:10,
    width = 2.1in,
    samples = 200,
    xtick = \empty,
    ytick = \empty,
    ]

    \addplot[
    green!50!black, thick,
    ]
    {.1*(x-2)*(x-6)*(x+2)+3};
    \node at (2,3) [label=below left:$\pp$, circle, fill, inner sep=1pt] (p) {};
    \draw [->, black] (p) -- +(axis direction cs:1,-1.6) node [right] {$T$};
    \draw [->, black] (p) -- +(axis direction cs:1.6,1) node [above] {$N$};
  \end{axis}
\end{tikzpicture}

    \caption{The Frenet frame.}
    \label{fig-frenet}
  \end{minipage}
  \end{figure}
\begin{remark}[Moving frames] In \cite{FO99}, the map $\rho$ appearing in step (ii) above is called \emph{a moving frame}.  For (locally) free actions this map is (locally) equivariant: 
\beq\label{r-equiv}\rho(\alpha(\g,\pz))=\rho(\pz)\cdot \g^{-1}.\eeq
 The term \emph{moving frame} underscores  the geometric origins of the construction, going back to the classical frames of Frenet and Darboux.  \exr{ex-se2}  can serve to illustrate this relationship. The Frenet frame of an oriented planar curve consists of  the properly oriented unit tangent $T$  and unit normal $N$ attached to a point $\pp$ on the curve. These vectors form an orthonormal matrix, and  combined with the point $\pp=(x,y)\in {\fld R}^2 $, can been viewed as  an element of  the  group $SE_2$.  This  point of view gives us  a map $\tilde \rho \colon J^1\to SE_2$, $\tilde\rho\left(x,y,\jt y1\right)=\left( [T,N], (x,y)\right)$, pictured in Figure~\ref{fig-frenet},
where 
$$ T=\begin{bmatrix}\frac 1{\sqrt{1+{\jt y1}^2}}\\ \frac {\jt y1}{\sqrt{1+{\jt y1}^2}}\end{bmatrix} \text{ and } N=\begin{bmatrix} -\frac {\jt y1}{\sqrt{1+{\jt y1}^2}}\\ \frac 1{\sqrt{1+{\jt y1}^2}}\end{bmatrix}.$$
The map $\rho$ defined by formulas \eqref{se2-rho}  (with the top signs)  in \exr{ex-se2} and the classical map $\tilde \rho$ above are related by  the inversion:
$$ \tilde\rho(\pz)= \rho(\pz)^{-1}\text{  for all  } \pz\in J^1.$$ 
The classical moving frame enjoys the left-equivariant property $\tilde\rho(\act(\g,\pz))=\g\cdot\tilde\rho(\pz)$ and is called a \emph{left} moving frame, while the map $\rho$ from \exr{ex-se2} has the right-equivariant property \eqref{r-equiv}, and is called a \emph{right} moving frame.
\end{remark}
Locality of the cross-section (and, therefore, of invariants)  and  non-constructiveness  of the step (\ref{st-solve})  are the main   drawbacks of Method~\ref{sm-method}. Nonetheless, it turned out to be very powerful  both for  explicit computation, and for  the study of  theoretical structural questions, in particular, in the studies of differential invariant algebra, and  the symmetry reduction of differential and difference equations and variational problems (see, for instance,  \cite{ko03:acta, olver11:diff-alg, mansf19}).

\subsection{Algebraic setting}\label{sect-cs-alg}
 Algebraic formulation of the cross-section method  turns it into a fully algorithmic procedure, 
  where the ``solving for the group element'' and ``substitution into the action" steps in Method~\ref{sm-method} are  replaced with the computation of  an elimination  ideal. The algebraic  formulation is applicable to rational actions of affine algebraic groups  on affine algebraic varieties and yields  generating sets of rational invariants defined on Zariski open subsets.\footnote{Unlike  open sets in metric topology, Zariski open subsets are dense in the ambient irreducible variety, and, therefore, the algebraic method is global.}   
 
 Throughout this section, $\fld K$ is a field of characteristic zero  and $\overline {\fld K}$ is its algebraic closure.\footnote{We take $\fld K=\fld R$ when drawing analogy  with the smooth setting,  while    we assume $\fld K=\fld Q$ when running computational  algorithms, and, in both cases, $\cfld K=\fld C$.} If $\mathcal V$ denotes  a variety then $\fld K [\mathcal V]$ denotes the ring of regular (polynomial) functions on $\mathcal V$ and $\fld K (\mathcal V)$ denotes the field of rational functions on $\mathcal V$. For $\mathcal U\subset \mathcal V$,  $\overline {\mathcal U}$ denotes the Zariski closure of  $\mathcal U$ in $\mathcal V$.  
 
We assume that,    in \defr{def-act},  the  group $\gva$ is a  set of $\fld K$-points of an affine algebraic group over $\cfld K$  defined by a polynomial ideal over $\fld K$.  In  more details, let $\wh \gva\subset{\cfld K}^\ell $ be the zero set of a radical equi-dimensional  ideal $G\subset \fld K[\gp]$, $\gp=(\gp_1,\dots,\gp_\ell )$ with  the group multiplication $m\colon \wh\gva\times\wh\gva\to  \wh\gva $ and the inversion map $i\colon \wh\gva\to \wh\gva$, being polynomial maps defined over $\fld K$. Let $\zva={\fld K}^n$ be an affine space. The action $\act$ of  $\gva$ on $\zva$ is a restriction of a rational action $\wh\act$ of $\wh\gva$  on $\wh\zva={\cfld K}^n,$ given, in coordinates, by $n$ rational functions over $\fld K$:
$$Z_i=\act_i(\gp,  z),\quad  i=1,\dots, n,$$
where both $z=(z_1,\dots, z_n)$ and $ Z=(Z_1,\dots,Z_n)$ are coordinate functions on $\wh\zva$, 
called the \emph{source} and the \emph{target} coordinates, respectively.  Two technical assumptions on the least common multiple $\delta$ of the denominators $\delta_i$'s of $\act_i$'s can be found in \cite[Assumption 2.2] {hk07:jsc}.

 Central to the construction of invariants is the set  $\ova$  consisting of pairs of points  belonging to the same orbit:
\beq\label{ova}\ova = \{(\pz,\apz)\in  \wh\zva\times\wh\zva \;|\;  \exists \g \in \wh\gva, \, \mbox{s.t.} \, \apz=\act(\g, \pz)\},\eeq
which, following \cite{PV89},  we call \emph{graph-of-the-action} set.

The ideal $\oid$ of the closure $\ov\ova\subset \wh\zva\times\wh\zva$ can be explicitly obtained as follows. 
We define  the ideal of the action in the polynomial ring  $\fld K[\gp,\mu, z,Z]$: 
\beq\label{i-act} A=\left<\left\{\delta_i (\gp,z)Z_i- \delta_i(\gp,z)\,\alpha_i(\gp,z)\right\}_{i=1}^n, \delta(\gp,z) \mu-1\right>+G\,\eeq
where $\mu$ is an additional variable introduced to ensure that the denominators are not zero. The ideal $\oid$ is obtained by elimination:
\beq\label{i-graph} O=A\cap\fld K[z,Z],\eeq
and we mainly use its extension $\oid^e\subset\fld K(z)[Z]$ in the ring of polynomial functions in $Z$ over the field of rational functions in $z$.\footnote{Any polynomial in $\oid$ can be viewed as a polynomial in the ring $\fld K(z)[Z]$. A generating set of $\oid$ generates $\oid^e$ in this larger ring.}  For a generic\footnote{We say that a property holds for a \emph{generic} $\pz$ in $\zva$ if there exists a Zariski open (and, therefore, dense) subset $\opsz\subset\zva$, such that the property holds for all $\pz\in\opsz$. Similarly, by \emph{generic orbits} we mean  orbits of points in a Zariski open subset $\opsz\subset\zva$.} $\pz$,  the specialization $O_{\pz}\subset \fld K[Z]$ of $O^e$
 is the ideal of the closure $\ov{\ova_{\pz}}$ of the orbit  of $\pz$.  Since  $\ova_{\g\pz}=\ova_{\pz}$ for any $\pz\in \zva$ and $\g\in \gva$,  we can expect  the coefficients of a canonical generating set of $\oid^e$  to be  invariant   under the action of $\gva$. Indeed, it is proven in   \cite[Section 2.3]{hk07:jsc}  that coefficients of the reduced \grob of   $\oid^e$ relative to any monomial order on $\fld K[Z]$ are rational invariants.  Moreover, the set of these  coefficients \emph{generates}  the field $\fld K(\zva)^\gva$  of rational invariants.  A simple algorithm for expressing any rational invariant $f(z)=\frac {p(z)}{q(z)}$ by computing   certain  normal forms of polynomial $p(z$) and $q(z)$ is presented in   \cite[Theorem 2.16]{hk07:jsc}.\footnote{ For a detailed comparison with previous related works by  Rosenlicht \cite{rosenlicht56}, M{\"u}ller-Quade and Beth \cite{beth99},  and M{\"u}ller-Quade and Steinwandt  \cite{quade99} see    \cite[Section 2.3]{hk07:jsc}. For a subsequent variation by Kemper see  \cite{kemper2016}.}
 \begin{example}[scaling] \label{ex-sc}Consider an action of   $\,\fld R^*$ of non-zero real numbers on $\fld R^2$ by scaling: $\alpha(a,x,y)=(a x, a y)$.
Then the group is described by an ideal $G=\left<a b-1\right>\subset \fld R[a,b]$, and the action ideal is: $$ A=\left<X-ax,\,\, Y-ay,\,\, ab-1\right>.$$
We compute the reduced \grob  of $A$  relative to the lexicographical\footnote{ A more efficient ``lexdegree'' elimination order is used in our Maple implementation.} monomial order with  $X<Y<x<y<a<b$: 
$$\{ \,yX - xY, \,\,ax - X,\, a y - Y, \,\,bX - x, \,\,bY- y, \,\,ab - 1\}.$$
The corresponding reduced \grob  of $\oid$ is $\{ yX - xY\}$, while  the  reduced \grob  of $O^e$ relative to the lexicographical monomial order with $X<Y$ is
\beq\label{gb-scaling}\left\{ Y - \frac   y x\,X\right\}.\eeq  
The only non-constant coefficient $h(x,y)=\frac y x$ is a generating rational invariant which we already met in \exr{ex-inv}. For an illustration of rewriting procedure, based on normal form computation, see Example~{2.17} in \cite{{hk07:jsc}}.
\end{example}
We enhance  the above construction by combining  the  graph-of-the-action ideal  with the cross-section ideal. We give an  algebraic counterpart\footnote{This definition is slightly different from the definition of a cross-section in \cite{hk07:jsc, hk07:focm} and is closer in spirit to the smooth definition. For algebraically closed fields, it coincides with definition of \emph{quasi-section} in Section 2.5 of \cite{PV89} and  of \emph{section} in \cite{hubert2013}. See  \cite{hubert2013} for a  comparison of these notions.} of  smooth \defr{def-cs}. %
\begin{definition}[Algebraic cross-section]\label{alg-cs} A subvariety $\cs\subset\zva$  defined by an ideal $K\subset \fld K[\zva]$ is a cross-section of degree $d$ for an action $\act\colon\gva\times\zva\to\zva$   if the corresponding subvariety $\wh\cs\subset\wh\zva$ defined by extending $K$ to $\cfld K[\wh\zva]$ is irreducible and intersects each generic orbit of the action $\wh\act\colon\wh\gva\times\wh\zva\to\wh\zva$ at exactly $d$ points. If $d=1$, the cross-section is called \emph{rational}.
\end{definition}
Let us consider an ideal $B=A+K\subset K[\gp,\mu, z,Z]$, where $A$ is defined by \eqref{i-act} with $K\subset\fld K[Z]$ written in  terms of target  coordinates. Then $(\g,\pz,\apz)\in \wh \gva\times \wh\zva\times\wh\cs$   is in the zero set of $B$ if and only if $\apz=\wh\act(\g,\pz).$ That means that the \emph{graph-section} set
\beq\label{iva}\iva= \{(\pz,\apz)\in\wh\zva\times \wh \cs \;|\;  \exists \g \in \wh\gva, \, \mbox{s.t.} \, \apz=\act(\g, \pz)\} \eeq
is the projection of the variety of $B$ to $\wh\zva\times \wh \zva$.  Its closure $\overline \iva$  is defined by the elimination ideal 
\beq\label{i-graph} I=B\cap\fld K[z,Z],\eeq
and we again use its extension $I^e\subset\fld K(z)[Z]$ in the ring of polynomial functions in $Z$ over the field of rational functions in $z$.   For a generic $\pz$,  the specialization $I_{\pz}\subset \fld K[Z]$ of $I^e$  is the ideal of the closure of  the intersection of  the orbit $\wh\ova_{\pz}$ with  $\wh\cs$. Therefore,   if $I_e$ is a zero-dimensional radical ideal and $\cfld K(z)[Z]\backslash I^e$  is a $d$-dimensional vector space, the ideal $K$ defines a cross-section of degree $d$.  As in the smooth case, a generic irreducible variety of codimension $\dimo$ (the maximal orbit dimension) is a cross-section, and, more specifically, a generic affine subspace of   codimension $\dimo$ is a cross-section.

As in the case of the ideal $\oid$, since  $\iva_{\g\pz}=\iva_{\pz}$ for any $\pz\in \zva$ and $\g\in \gva$, we expect the coefficients of a reduced \grob of $I^e$   to be invariant  under the action $\act$. 
The  proof of the following theorem can be found in   \cite[Section 3.2]{hk07:jsc} and \cite{hubert2013}.
\begin{theorem}[Generators of $\fld K(\zva)^\gva$] \label{thm-alg}  For an action $\act\colon\gva\times\zva\to\zva$ and a cross-section $\cs$, the coefficients of the reduced Gr\"obner basis  $Q$ of the ideal $I^e\subset\fld K(z)[Z]$, relative to any monomial order on the variables $Z$, is a generating set of  the field   $\fld K(\zva)^\gva$ of rational invariants.

 If  the cross-section  is rational, then 
$Q=\left\{Z_i-r_i(z)\right\}_{i=1}^n$,  and  the rational  invariants $r_1,\dots, r_n$ have the replacement property: 
$$f(z)=f(r_1,\dots, r_n), \text{ for any  }f(z)\in \fld K(\zva)^\gva.$$
\end{theorem} 
The first  statement of the  theorem leads to a practical algorithm for computing a generating set of rational invariants, while the last statement  asserts  that a \emph{rational cross-section} yields a generating  set of rational invariants with the same replacement properties as a set of smooth normalized invariants, and  so they can be used to define the process of invariantization. There are, however,  actions  for which  rational cross-sections do not  exist.\footnote{In Section 2.5 of \cite{PV89}, what we call a \emph{rational cross-section} is called just a \emph{section}, and a necessary and sufficient condition of  its existence is discussed.} Theorem 2.17 of \cite{hk07:focm} (see also \cite[Section 2.5]{PV89}), shows that  for a cross-section $\cs$ of degree $d$, the  field   $\fld K(\cs)$ is an algebraic extension of  degree $d$ of the invariant field ${\fld K(\zva)^\gva}$, and so for rational cross-sections ${\fld K(\zva)^\gva}\cong \fld K(\cs)$.  For a cross-section of degree $d$, there exists $d$ tuples of $n$ invariants  in the algebraic closure  $\overline{\fld K(\zva)^\gva}$ with the replacement  properties. An algebraic invariantization process can be defined in terms of these invariants as shown in   \cite[Section 2.6]{hk07:focm}.

 \begin{example}[scaling revisited]\label{ex-sc-cont} We revisit \exr{ex-sc} and, first, consider a rational cross-section  defined by an ideal $K_1=\ideal{X-1}$. Then
 $$B_1=A+K_1=\ideal{X-1,\,\,X-ax, \,\,Y-ay,\,\, ab-1}$$
 and
 $$ I^e_1=\ideal{X-1, Y-\frac y x}.$$
 The replacement  invariants are 
 $$r_1=1 \text{ and } r_2=\frac y x.$$
 If $f(x,y)$ is any rational invariant, then $f(x,y)=f\left(1,\frac yx\right).$ 
 
Secondly, we consider a  cross-section of degree 2  defined by an ideal $K_2=\ideal{X^2+Y^2-1}$. 
 Then
 $$B_2=A+K_2=\ideal{X^2+Y^2-1,\,\,X-ax, \,\,Y-ay, \,\, ab-1}$$
 and
 $$ I^e_2=\ideal{  Y-\frac  y x X,\,\,X^2-\frac {x^2}{x^2+y^2}}.$$

There are two zeros of $ I^e_2$ over $\overline{\fld R(\zva)^\gva}:$
$$ \xi_1=\left(\frac {x}{\sqrt{x^2+y^2}},\,\frac {y}{\sqrt{x^2+y^2}}\right) \text{ and  }  \xi_2=\left(-\frac {x}{\sqrt{x^2+y^2}},\,-\frac {y}{\sqrt{x^2+y^2}}\right).$$
Components of each of these zeros have the same replacement properties as smooth normalized invariants in Section~\ref{sect-cs-sm}.
\end{example}
Computational advantage of using ideal $I^e$ instead of $O^e$ for finding a generating set of the field $\fld K(\zva)^\gva$ will be apparent in \exr{ex-bf} below. 
  In our next example, we underscore why,  in \defr{alg-cs} of the cross-section, it is important to use the varieties over $\cfld K$ rather than over $\fld K$.
   \begin{example}[scaling modified]\label{ex-scR} We modify our scaling \exr{ex-sc} by considering the action 
   \beq\label{sc2}\tilde\act(a,x,y)=(a^2 x, a^2 y)\eeq
   of   $\gva=\fld R^*$ on $\zva=\fld R^2$.
Note that for  the action of $\wh \gva =\fld C^*$ on $\wh\zva =\fld C^2$ defined by \eqref{sc2}, the orbits are the same as for the action in \exr{ex-sc}, namely  the punctured  lines through the origin, with the origin removed, and the origin itself. Thus, in accordance with \defr{alg-cs}, $K_1$ and $K_2$ in \exr{ex-sc-cont} define two cross-sections of degree 1 and 2, respectively for the modified action $\tilde\act$ as well.
Ideals $O^e$ and $I^e$, appearing in  Examples~\ref{ex-sc} and~\ref{ex-sc-cont},  would not change either, and the computation of  generating invariants in these examples  remains valid.

However, if, in \defr{alg-cs}, we did not introduce the variety $\wh\cs\subset \wh\zva$    over  $\cfld K$ and just  said that  the variety $\cs\subset \zva$  of $K$ is a cross-section of degree $d$ if it intersects generic orbits in  $\zva$  at $d$ simple points, then the variety $\cs_1\subset \fld R^2$ of $K_1$ would not satisfy the definition  of cross-section, while the variety $\cs_2\subset \fld R^2$ of $K_2$ would be a cross-section of degree 1.
Indeed,  in $\fld R^2$ under  the action \eqref{sc2},  two opposite open rays become distinct orbits (as for the smooth action $\act_2$ in \exr{ex-act-orb}). Therefore,   there is no Zariski open subset on which $\cs_1$ intersects each orbit, while  $\cs_2$ intersects each  generic orbit only once, and, therefore, would  have degree 1, making the second statement of \ther{thm-alg}  invalid.  
\end{example}

  \exr{ex-scR} can be also used to underscore the difference in separating properties of  generating sets  over algebraically closed and non-closed fields. \emph{Over an algebraically closed field}, it is known that \emph{a set of rational invariants separates generic orbits if and only if this set generates the field of rational invariants}.\footnote{See, for instance,  \cite[ Lemma 2.1 and Theorem 2.3]{PV89}.}
 As we see from \exr{ex-scR}, this statement is not true over $\fld R$.    Invariant $r_2=\frac y x$ generates the field of rational invariants under the action \eqref{sc2} on $\fld R^2$, but it takes the same value on any two opposite rays, which are two different  orbits for this action. Therefore, there is no Zariski open subset of $\fld R^2$ on which $r_2$ separates orbits.  
 The converse statement  is also not true over $\fld R$. Indeed, for the action $\act_1$ in \exr{ex-act-orb}, $f(x,y)=y^3$ is a separating invariant, but is not generating.  
      
  We conclude this section with an example from the classical invariant theory.
      \begin{example}[Binary forms] \label{ex-bf}  Any action $\act_1$ of a group $\gva$ on ${\fld K}^n$ induces an action $\act_2$  on the set of functions on ${\fld K^n}$. For  $\g\in \gva$ and a function  $f(\pz),\, \pz\in\fld K^n$, we set $\act_2(\g,\, f)=F$,  where\footnote{The appearance of $\g^{-1}$, rather than $\g$ in the formula above, is necessary  to ensure that   property (i) in \defr{def-act} is satisfied.}  
\beq\label{act-f}F (\pz)=f\left(\act(\g^{-1}, \pz)\right).\eeq

In this example, we start with the standard action $\act_1$ of $GL_2(\fld C)$  (or of its subgroup  $SL_2(\fld C)$)  on $\fld C^2$:       $$X=a_{11} x+a_{12} y, \quad Y=a_{21} x+a_{22} y$$
and then consider the induced action $\act_2$ on the linear space $\bin_m$  of  binary forms  of degree $m$:
$$ \displaystyle{f_m(x,y)=\sum_{i=0}^m\binom{m} { i} c_ix^{i} y^{m-i}}.$$
The action \eqref{act-f}  can be described by the action on the coefficients $c_0,\dots, c_m$. For instance,  a binary quadratics  $f(x,y)=c_2x^2+2c_1xy+c_0y^2$  is transformed to   $F(x,y)=C_2x^2+2C_1xy+C_0y^2$, where, with $\delta=a_{11} a_{22}-a_{12} a_{21}$, we have:
\begin{align*}
C_2&=\frac 1 {\delta^2}\left({a_{21}}^2c_0-2a_{21}a_{22}c_1+{a_{22}}^2c_2\right),\\
C_1&=\frac 1 {\delta^2}\left(-a_{11}a_{21}c_0+(a_{11}a_{22}+a_{12}a_{21})c_1-a_{12}a_{22}c_2\right),\\
C_0&=\frac 1 {\delta^2}\left({a_{11}}^2c_0-2a_{11}a_{12}c_1+{a_{12}}^2c_2\right).
\end{align*}
In the classical invariant theory,  invariant functions under the action $\act_2$ (they depend only on the coefficients) are  called \emph{invariants}, while invariant functions under the product action $\act=\act_1\times \act_2$ on $\fld C^2\times \bin_m$  (they depend  both on the coefficients and the variables $(x,y)$) are called \emph{covariants}.
We invite the reader to verify that a binary form $f_m(x,y)$ is a $GL_2$-covariant. On the space $\bin_2$ of binary quadratic,  under the $GL_2$-action, there are no non-constant rational invariants and $f_2(x,y)$ generates  the field of rational covariants.  Under the $SL_2$-action, the field of rational invariants is generated by the discriminant $\Delta_2={c_1}^2-c_0c_2$, while the field of rational covariants is generated by $f_2(x,y)$ and $\Delta_2$. 
For $m>2$, it  is known  (see \cite[Theorem~2.12]{PV89}) that  for the $SL_2$-action   on $\bin_m$, the field of rational invariants $\fld{C}(\bin_m)^{SL_2}$ is generated by $m-2$ \emph{algebraically independent} invariants, while the field of rational covariants $\fld C\left(\fld C^2\times \bin_m\right)^{SL_2}$  is generated by $m$ \emph{algebraically independent} covariants. For the $GL_2$-invariants and covariants, there are similar results with the number of  algebraically independent generators reduced by one.\footnote{This result is in  stark contrast  with the results  about the minimal numbers of generators for the algebras of \emph{polynomial} invariants and covariants. In \cite{LercierOlive2017}, the authors compute a minimal set of 476 generators (resp. 510 generators) for the algebra of  polynomial $SL_2$-covariants for $\bin_9$ (resp. $\bin_{10}$).  The references for lower-degree results can be found in \cite[Table 1]{LercierOlive2017}. To the best of my knowledge,  the explicit results for $m>10$ are unknown.}   In principle, these rational invariants and covariants can be computed by the algebraic cross-section method presented in this section. In practice, as the degree $m$ grows, the Gr\"obner basis computation becomes unmanageable.\footnote{Independent generators of the field of \emph{rational} invariants for binary octavics, $\bin_8$, obtained in \cite{maeda1990}, along with some interesting structural results for general $m$, demonstrate the complexity of the expressions.} We invite  the reader to play with different degrees of forms and cross-sections using the {\sc Maple} code \cite{suppl} and, perhaps, to write a more efficient code to compute rational invariants and covariants of binary forms.
Here we included a few snapshots.

 For the $SL_2$-action on the space $\bin_3$ of binary cubics with the cross-section defined by $C_0=0$, we obtain the graph-section  ideal
$$ I^e=\ideal {C_0,\,\, -4 {C_1}^{3} {C_3} +3 {C_1}^{2}{C_2}^{2}+\Delta_3},$$
where $\Delta_3=(c_0c_3-c_1c_2)^2-4(c_1c_3-{c_2}^2)(c_0c_2-{c_1}^2)$
is the discriminant of the binary cubic $f_3$. For the $GL_2$-action on $\bin_3$ there are no invariants, but two independent generating covariants. The graph-section ideal  with the cross-section ideal  $K=\ideal{X-1,\,Y,C_0-1,\,\,C_1}$ is
$$I^e=K+\ideal{C_3-f_3(x,y), \,\,{C_2}^6+4\,\frac{{H_3}^3}{ \Delta_3} {C_2}^3+\frac{{f_3}^2{H_3}^3}{ \Delta_3}},$$
where $H_3=(c_1c_3-{c_2}^2)x^2-(c_0c_3-c_1c_2)xy+(c_0c_2-{c_1}^2)y^2$ 
is $1/36$ of the Hessian of $f_3$.  From the coefficients of the $I^e$-generators, we see that $f_3$ and  $\frac {{H_3}^3}{\Delta_3}$ comprise a minimal  generating set of the field of rational covariants: $\fld C\left(\fld C^2\times \bin_2\right)^{GL_2}\cong \fld C\left(f_3,\frac {{H_3}^3}{\Delta_3} \right)$. 

 \end{example}
 
\section{Differential Signatures}\label{sect-sig}
\exr{ex-bf}, devoted to  binary forms,  underscores that although a minimal generating set of rational invariants, in principle,  can be found  by a computational  algebra  algorithm, in practice, this is not  feasible as the number and complexity of  invariants increases  with the degree of the form. 
 In this section, we present an alternative method,  based on \emph{differential invariants},  for solving the equivalence problem for binary forms, as well as many other equivalence problems. The \emph{differential signature} construction \cite{calabi98}  originated from  Cartan's equivalence method  \cite{C53}.\footnote{ For a modern exposition see \cite[Chapter 8]{olver:purple}, in particular, the notion of classifying manifolds.}
For  curves in $\fld R^2$ or $\fld C^2$, under either effective-on-all-open-subsets  smooth local actions of Lie groups or effective rational actions of algebraic groups, it consists of the following steps:
 \begin{method}\emph{({Differential signatures of planar curves})}\label{method-sig}
 \begin{enumerate}[i.]
 \item \emph{(Prolong the action)} 
 An action of a real (or complex) group $\gva$ on  $\fld R^2$ (or $\fld C^2$)  is prolonged to the jet space  $J^r$ of  curves, where $r=\dim\gva$, using formulas \eqref{pr-act}.
 \item \emph{(Construct  a pair of classifying invariants)}  In the smooth setting, on an open subset of $J^r$, there exists  a pair of smooth functionally independent invariants  \cite[Theorem 5.24]{olver:purple}, while in the algebraic setting there exists a pair of  rational algebraically  independent generating invariants  \cite[Theorem~2.20]{krv20:siaga}.  In both settings, we will call such a pair \emph{classifying}. 
 \item \emph{(Construct a signature map)}  A pair of classifying invariants  can be  \emph{restricted} to a given planar curve $\Gamma$. As the result, we obtain a  signature map, which sends $\Gamma$ to another planar curve $\sig_\Gamma$, called the \emph{signature}\footnote{For most actions, there is a very small class of curves, called exceptional curves, for which their signatures are not defined. For some curves, their signatures degenerate to a point. For example, the Euclidean signature of a circle is a point. Such degeneration indicates that  the symmetry subgroup of a curve  has a positive dimension.}  of  $\Gamma$.    Since the signature  is based  on invariants,  two $\gva$-equivalent curves have the same signature. The converse is true for analytic or algebraic curves. For smooth  non-analytic curves the situation is more subtle  as demonstrated in \cite{Musso2009, Hickman2012, olver15, GK-2021}.  
\end{enumerate}
\end{method}

\begin{figure}
    \hspace{-1cm}
  \begin{subfigure}[b]{0.5\linewidth}
  \centering
    \includegraphics[width=0.9\linewidth]{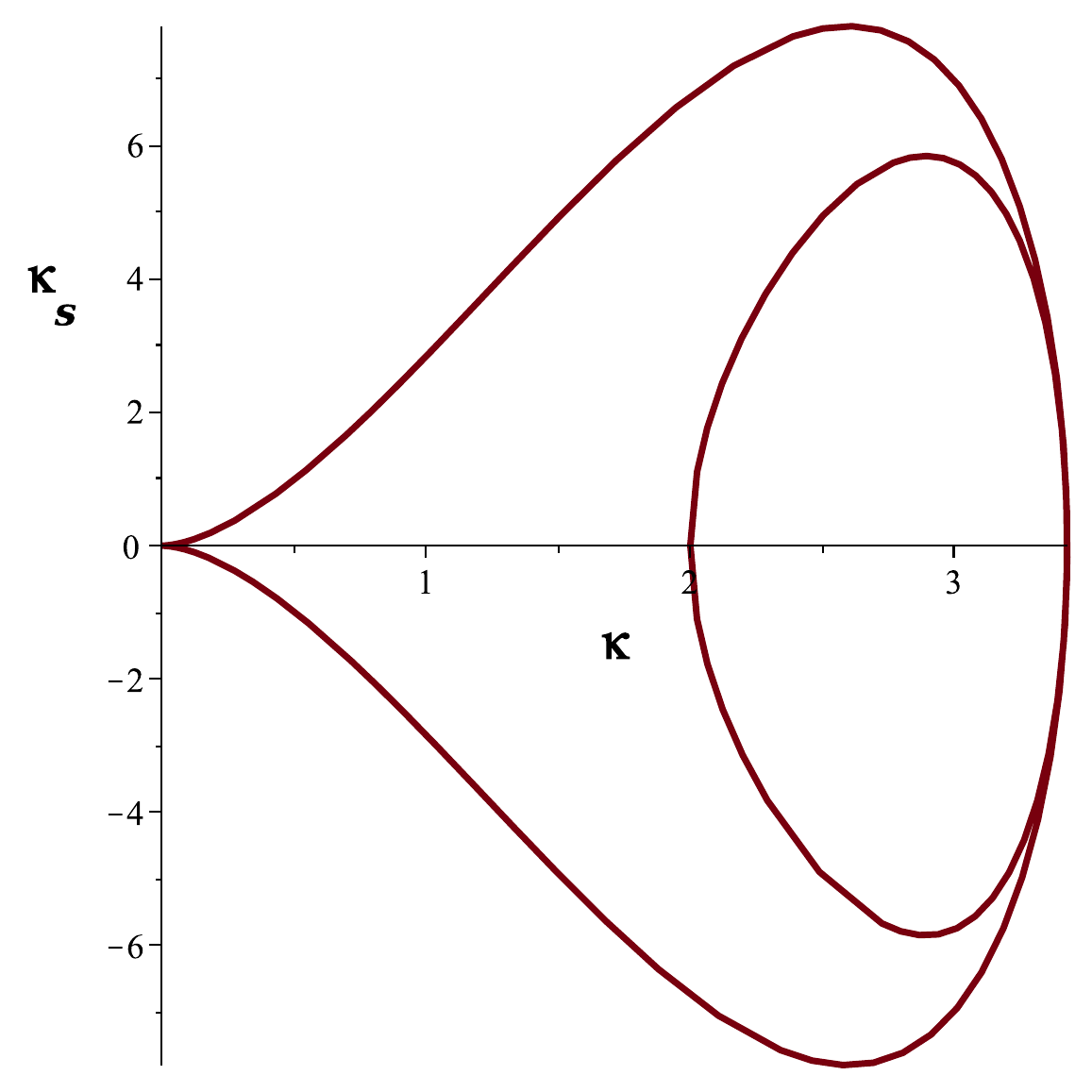}
     \caption{Signature of $\Gamma_1$.}
     \label{fig:siga}
  \end{subfigure}
  %
  \begin{subfigure}[b]{0.5\linewidth}
   \centering
    \includegraphics[width=0.9\linewidth]{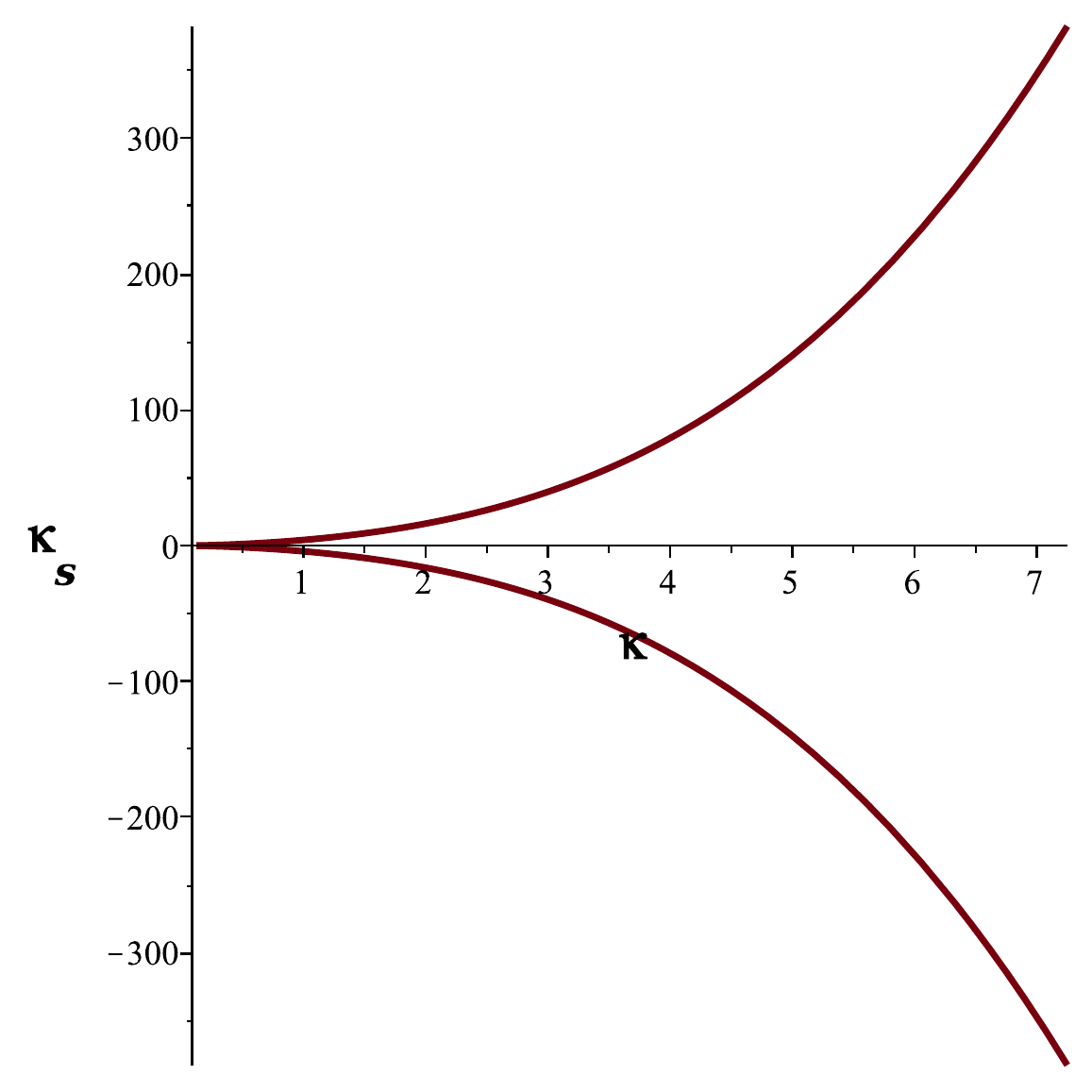}
     \caption{Signature of $\Gamma_2$.}
      \label{fig:sigb}
  \end{subfigure}
\caption{$SE_2$-signatures of curves $\Gamma_1$ and $\Gamma_2$ in \exr{euc-sig}.}
 \Description{Two pictures showing two different planar curves.}
\label{fig:sig}
 \end{figure}

 \begin{example}[Euclidean signature]\label{euc-sig} In \exr{ex-se2}, we prolonged the standard $SE_2$-action on the plane to the second jet space $J^2$ and used the cross-section method to find a lowest order differential invariant, the Euclidean curvature,  $\inv \jt y2=\kappa$. We note that $\dim SE_2=3$  and  leave it to the reader to prolong  the action to $J^3$ using formulas \eqref{pr-act}. Equations  \eqref{cs-se2}, used to define a cross-section on $J^2$, define a cross-section on $J^3$ as well,  and substitution of  the moving frame map $\rho$ defined by \eqref{se2-rho} (with appropriate signs) into $\jt Y 3$ yields a third order  normalized invariant, where for better readability, we replaced the jet variables with the  corresponding derivatives:
\beq\label{kappas}\inv \jt y 3=\frac {\left(1+y_x^2\right)y_{xxx}-3\, y_x\,y_{xx}^2}{\left(1+y_x^2\right)^3}.\eeq
 This invariant is, in fact, the derivative of the curvature $\kappa$ with respect to the arc-length, often denoted as $\kappa_s$.  Thus we obtain a classifying pair of invariants $(\kappa,\kappa_s)$.
 
We will construct Euclidean signatures for two non-congruent rational cubics.\footnote{Signature construction for  rational and implicit algebraic curves based on \emph{rational} classifying invariants can be found in  \cite{bkh13:sigma, krv20:siaga}.}
To construct the signature of  a rational  cubic $\Gamma_1$  with  parameterization
\beq\label{Gamma1} x(t)=t^2-1,\quad y(t)=t(t^2-1)    
\eeq
we need to restrict the classifying pair $(\kappa,\kappa_s)$ to $\Gamma_1$. To this end, we \emph{lift} $\Gamma_1$ to $\jt{\Gamma_1} 3\subset J^3$ with  parameterization:
\begin{align*}
\jt y1(t) =&\frac{\dd y t}{\dd x t }=\frac{3t^2-1}{2t}, \quad \jt y 2(t) =\frac{\dd {\jt y 1} t}{\dd x t }=\frac{3t^2+1}{4t^3},\\
 \jt y 3(t)=&\frac{\dd {\jt y 2} t}{\dd x t }=-\frac{3t^2+3}{8t^5},
 \end{align*}
  and substitute $y_x=\jt y1(t), \, y_{xx}=\jt y2(t), \, y_{xxx} = \jt y3(t)$ into  formula \eqref{kappa} for $\kappa$ and  formula \eqref{kappas} for $\kappa_s$: 
\beq\label{sig1}\kappa|_{\Gamma_1}=\frac{6t^2+2}{(9t^4-2t^2+1)^{3/2}}\text{ and } \kappa_s|_{\Gamma_1}=-24 t\frac{9t^4+4t^2-1}{(9t^4-2t^2+1)^{3}}. \eeq
The signature  of $\Gamma_1$ is the curve parameterized  by \eqref{sig1}, plotted in Figure~\Ref{fig:siga}.

Let us consider a rational  cubic $\Gamma_2$  with  parameterization
\beq\label{Gamma2} x(t)=t^2,\quad y(t)=t^3.  \eeq
  Its signature curve is parameterized by 
  \beq\label{sig2}\kappa|_{\Gamma_2}=\frac{6}{\sqrt{t^2(9t^2+4)^3}}\text{ and } \kappa_s|_{\Gamma_2}=-24\frac{9t^2+1}{(9t^2+4)^3t^{3}}. \eeq
  and is shown in Figure~\ref{fig:sigb}. 
   
  We observe that the signatures of two $SE_2$-non-equivalent curves $\Gamma_1$ and $\Gamma_2$ are different, and invite the reader  to check that any curve obtained  from $\Gamma_1$ by rotation or translation has the same signature as $\Gamma_1$ (and similarly for $\Gamma_2$).  
 \end{example}

\begin{figure}
  \centering
     \begin{tikzpicture}[scale=1]
  \begin{axis}[
    axis lines = center,
    xlabel = {$x$},
    ylabel =  {$y$},
    zlabel = {$y^{(r)}$},
    xmin = -3, xmax = 5,
    ymin = -25, ymax = 25,
    zmin = -5, zmax = 5,
    plot box ratio = 2 8 1,
    width = 5in,
    samples = 50,
    samples y = 1,
    xtick = \empty,
    ytick = \empty,
    ztick = \empty,
    view = {5}{5},
    ]

    \addplot3[
    green!60!black,
    thick,
    domain = -2:4,
    ]
    (x,{x*(x+1)*(x-4)+1},{.2*(3*x^2-6*x-4)})
    node[below right,pos=.9] {\large$\Gamma^{(r)}$};

    \addplot3[
    green!30!black,
    thick,
    domain = -2:4,
    ]
    (x,{x*(x+1)*(x-4)+1},0)
    node[below right,pos=.9] {\large$\Gamma$};

    \addplot3[
    red,
    thick,
    domain = -2:4,
    ]
    (3,{x*(x+1)*(x-4)+1},{.2*(3*x^2-6*x-4)})
    node[below right,pos=.9] {\large$\mathcal S_\Gamma$};

    \fill[blue, opacity=.2] (3,-15,-4) -- (3,-15,5) -- (3,5,5) -- (3,5,-4) -- cycle;
    \node [blue] at (3.3,-12,-2) {\large $\mathcal K$};

    \foreach\z in {-4,...,4}
    \addplot3[
   purple!80!black, 
    dashed,
    samples = 20,
    domain = -3:5,
    ]
    (x, 0, {-.01*x^2+\z});
  \end{axis}  
\end{tikzpicture}
   \caption{A signature as a projection onto a cross-section.} 
   \Description{A planar curves, a space curve, a surface, projection of a space to the surface along the curved lines. }
\label{fig:sig-geom}
 \end{figure}
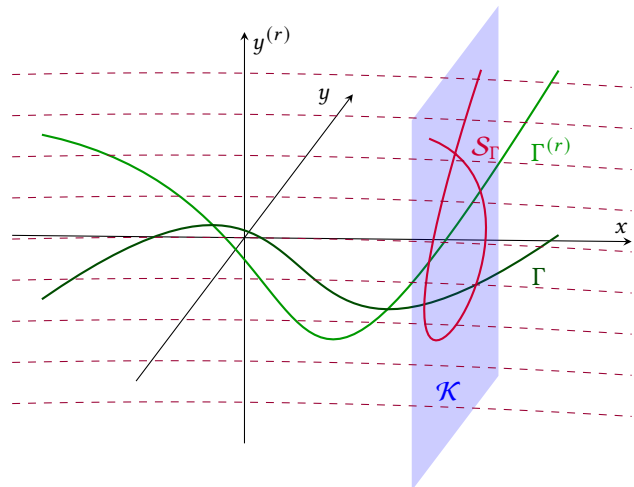

Signatures  based on normalized invariants\footnote{Such as $(\kappa, \kappa_s)$-signature in \exr{euc-sig}.}  constructed using a cross-section $\cs\subset J^k$ have a simple geometric interpretation presented in Figure~\ref{fig:sig-geom}: first, lift a curve to  the jet bundle $J^k$ and then  project it to $\cs$ along the orbits.  
  The signature is the image of this projection. 
  
  Despite this simple interpretation, an algorithmic solutions for deciding equality  of two signatures in the smooth setting is challenging.  In the algebraic setting, the normalized invariants are rational if and only if the cross-section is rational. Nonetheless, as discussed in Method~\ref{method-sig}, a classifying pair of rational invariants always exists, and so, in theory (but often not in practice),   one can  compute  an implicit equation of a signature. In \cite{krv20:siaga}, it is shown how to determine the degree of a signature curve, without computing  the  signature itself, while in \cite{DuffRuddy2023} numeric methods for signature comparison are proposed.
  We now revisit \exr{ex-bf} and show how  signatures of binary forms can be constructed, following \cite[Chapter 8]{olver:inv}.
  \begin{example}[Signatures of binary forms] \label{ex-bfs} To construct a signature of a binary form we  first turn it into a polynomial in one variable, converting
  $f_m(x,y)$ into $\phi_m(p)=f_m(p,1)$. Then $f_m(x,y)=y^m\phi_m\left(\frac x y\right)$. We now can identify  a binary form  of degree $m$ with a graph of function $q=\phi_m(p)$. The $GL_2$-action on the space of binary forms  $\bin_m$ induces the following action on the  $(p,q)$-plane: 
     $$P=\frac{a_{11}p+a_{12}}{a_{21}p+a_{22}}, \quad Q=\frac q{(a_{21}p+a_{22})^m}$$
and an action on non-homogenous polynomials of degree $m$ sending   $\phi_m(p)$ to a polynomial $\Phi_m(p)=(a_{21}p+a_{22})^m\phi\left(\frac{a_{11}p+a_{12}}{a_{21}p+a_{22}}\right)$.
     As it is shown in \cite{olver:inv}, the signature based on  a pair of classifying  invariants $(T_1,T_2)=\left(\frac {T^2} {H^3},\frac {V} {H^2}\right)$ of differential orders 3 and 4, respectively\footnote{In fact, we choose a slightly different, but equivalent to the one in \cite{olver:inv}, pair of classifying invariants.}, where
     \begin{align*}
     H&=q\, \jt q 2-\frac {m-1} m \left(\jt q 1\right)^2
\end{align*}
is, up to a constant multiple, the inhomogeneous counterpart of the Hessian of $f_m$, and 
  \begin{align*}
     T&=q^2 \jt q 3-3\frac {m-2} m q \jt q 1 \jt q2 +2 \frac {(m-1)(m-2)} {m^2}\left(\jt q 1\right)^3,\\
    V &= q^3 \jt q 4 - 4 \frac{m-3}{m}\, q^2\jt q1 \jt q 3 +
6 \, \frac{(m-2)(m-3)}{m^2}\, q\,\left(\jt q 1\right)^2\jt q2\\
&-3 
\,\frac{(m-1)(m-2)(m-3)} {m^3}\left(\jt q 1\right)^4,
     \end{align*} 
     serves to solve the equivalence problem for binary forms, whose Hessian is non-zero. 

     Let us construct the signature of a binary quartic
    \beq\label{q1a}f(x,y)=x^4+y^4.\eeq Its inhomogeneous  counterpart is $\phi(p)=p^4+1$.
     The restrictions of classifying invariants to $\phi$
     \beq\label{par-1a}T_1|_{\phi}=\frac{\left(p^{4}-1\right)^{2}}{3 p^{4}},\quad T_2|_\phi=
\frac{p^{8}-p^{4}+1}{6 p^{4}}\eeq
parametrize the signature curve of \eqref{q1a}. The implicit equation of the signature:
\beq\label{sig-1}6T_2-3T_1-1=0.\eeq
 defines a straight line. Over $\fld C$, the image of parametrization \eqref{par-1a}  equals to this line, while over $\fld R$, the image is  a non-dense subset, namely a ray where $T_1\geq0$, while the opposite ray is the signature of the signature of
   \beq\label{q1b}\tilde f(x,y)=x^4-y^4.\eeq
   Quartics \eqref{q1a} and \eqref{q1b} are $GL_2(\fld C)$ equivalent,  but they are not $GL_2(\fld R)$-equivalent.\footnote{We invite the reader to trace the non-injective parametrization \eqref{par-1a} over $\fld R$ of the appropriate ray  and note that the number  of  real parameters $p$ in the  preimage of a  generic  point is four. How does this number change over $\fld C$? For the relationship between the cardinality of the symmetry group of a binary form  and  the cardinally of the preimage of a  generic  point under the signature map see \cite{bo00}.}  
     Let us now construct  the signature of  
   \beq\label{q2a} g(x,y)=x^4+x^2y^2+y^4.\eeq Its inhomogeneous  counterpart is $\psi(p)=p^4+p^2+1$.
     The restrictions of classifying invariants to $\psi$,
     \beq\label{par-2a}T_1|_{\psi}=\frac{324 p^{2} \left(p^{4}-1\right)^{2}}{\left(2 p^{4}+11 p^{2}+2\right)^{3}},\,\,T_2|_\psi=
\frac{3(16 p^{8}+20 p^{6}+9 p^{4}+20 p^{2}+16)}{2\left(2 p^{4}+11 p^{2}+2\right)^{2}},
\eeq
parametrize the signature curve of \eqref{q2a}, whose implicit equation  is a cubic
 \begin{align}\label{sig-2}9800 T_{2}^{3}&-19773 T_{1}^{2}+79092 T_{1} T_{2}-64392 T_{2}^{2}\\
\nonumber &-13182 T_{1}+33714 T_{2}-972=0.\end{align}
Over $\fld C$, the image of parametrization \eqref{par-2a} is dense in the zero set of \eqref{sig-2}, while over $\fld R$, the image is  a non-dense subset, where $T_1\geq0$, while the other ``half''  of the zero set is the signature of
   \beq\label{q2b}\tilde g(x,y)=x^4-x^2y^2+y^4.\eeq
The above examples underscore that, over $\fld C$,  the equality of  signatures can be decided by computing their implicit equations, while over $\fld R$ a more subtle analysis is necessary.
      \end{example}

 The crucial advantage of the signature method in comparison with separating invariants approach, for solving equivalence problems for $m$-ary forms, is that the same relatively small set of classifying invariants can be used  for the forms of all degrees.
The classifying set of invariants for ternary forms  was first obtained in \cite{Kthesis} and applied in \cite{km02} for ternary cubics. 
It is worth noting, that in addition to solving equivalence problems, signature maps can be used to obtain information about  global symmetry groups of objects, as well as  their groupoids of local symmetries  \cite{bo00, olver15, krv20:siaga, GK-2021}.

\bibliographystyle{ACM-Reference-Format}


\end{document}